\documentclass[conference]{IEEEtran}
\IEEEoverridecommandlockouts
\usepackage{cite}
\usepackage{url}
\makeatletter
\newcommand*{\rom}[1]{\expandafter\@slowromancap\romannumeral #1@}
\makeatother
\usepackage{amsmath,amssymb,amsfonts}
\usepackage{algorithmic}
\usepackage{graphicx}
\usepackage{textcomp}
\usepackage{xcolor}
\def\BibTeX{{\rm B\kern-.05em{\sc i\kern-.025em b}\kern-.08em
    T\kern-.1667em\lower.7ex\hbox{E}\kern-.125emX}}
\begin{document}

\title{Advancing Speech Recognition With No Speech Or With Noisy Speech\\
{
}
}

\author{\IEEEauthorblockN{Gautam Krishna}
\IEEEauthorblockA{\textit{Brain Machine Interface Lab} \\
\textit{The University of Texas at Austin}\\
Austin, Texas \\
}
\and
\IEEEauthorblockN{Co Tran}
\IEEEauthorblockA{\textit{Brain Machine Interface Lab} \\
\textit{The University of Texas at Austin}\\
Austin, Texas \\
}
\and
\IEEEauthorblockN{Mason Carnahan}
\IEEEauthorblockA{\textit{Brain Machine Interface Lab} \\
\textit{The University of Texas at Austin}\\
Austin, Texas \\
}
\and
\IEEEauthorblockN{Ahmed Tewfik}
\IEEEauthorblockA{\textit{Brain Machine Interface Lab} \\
\textit{The University of Texas at Austin}\\
Austin, Texas  \\
}
}

\maketitle

\begin{abstract}
In this paper we demonstrate end-to-end continuous speech recognition (CSR) using electroencephalography (EEG) signals with no speech signal as input. An attention model based automatic speech recognition (ASR) and connectionist temporal classification (CTC) based ASR systems were implemented for performing recognition. We further demonstrate CSR for noisy speech by fusing with EEG features.
\end{abstract}


\section{Introduction}
Electroencephalography (EEG) is a non invasive way of measuring electrical activity of human brain. In \cite{krishna2019speech} we demonstrated deep learning based automatic speech recognition (ASR) using EEG signals for a limited English vocabulary of four words and five vowels. In this paper we extend our work for a much larger English vocabulary and we use state-of-art end-to-end continuous speech recognition models to perform recognition. In our prior work we predicted isolated words and vowels.\\
ASR systems forms the front end or back end in many cutting edge voice activated technologies like Amazon Alexa, Apple Siri, Windows Cortana, Samsung Bixby etc. Unfortunately these systems are trained to recognize text only from acoustic features. This limits technology accessibility to people with speaking disabilities and disorders. The research work presented in this paper tries to address this issue by investigating speech recognition using only EEG signals with no acoustic input and also by combining EEG features along with traditional acoustic features to perform recognition. We believe the former will help with speech restoration for people who can not speak at all and the latter will help people who are having speaking disabilities like broken or discontinued speech etc to use voice activated technologies with better user experience there by helping in improving technology accessibility.\\
ASR performance is degraded in presence of noisy speech and in real life situations most of the speech is noisy. Inspired from the unique robustness to environmental artifacts exhibited by the human auditory cortex \cite{yang1991auditory,mesgarani2011speech} we used very noisy speech data for this work and demonstrated lower word error rate (WER) for smaller corpus using EEG features, concatenation of EEG features and acoustic features. \\

In \cite{wang2017simulation} authors decode imagined speech from EEG using synthetic EEG data and connectionist temporal classification (CTC) network but in our work we use real EEG data, use EEG data recorded along with acoustics. In \cite{kumar2018envisioned} authors perform envisioned speech recognition using random forest classifier but in our case we use end to end state of art models and perform recognition for noisy speech. In \cite{ramsey2017decoding} authors demonstrate speech recognition using electrocorticography (ECoG) signals, which are invasive in nature but in our work we use non invasive EEG signals. \\
This work is mainly motivated by the results explained in \cite{krishna2019speech,zhao2015classifying,sun2016neural,wang2017simulation}. In \cite{zhao2015classifying} the authors used classification approach for identifying phonological categories in imagined and silent speech but in our work we used continuous speech recognition state of art models and our models were predicting words, characters at each time step. Similarly in \cite{sun2016neural} neural network based classification approach was used for predicting phonemes.\\

Major contribution of this paper is the demonstration of end to end continuous noisy speech recognition using only EEG features and this paper further validates the concepts introduced in \cite{krishna2019speech} for a much larger English corpus.

\section{Automatic Speech Recognition System Models}
An end-to-end ASR model maps input feature vectors to an output sequence of vectors of posterior probabilities of tokens without using separate acoustic model, pronunciation model and language model.
In this work we implemented two different types of state of art end to end ASR models used for the task of continuous speech recognition and the input feature vectors can be  EEG features or concatenation of acoustic and EEG features. We used Google's tensorflow and keras deep learning libraries for building our ASR models.

\subsection{Connectionist Temporal Classification (CTC)}
The main ideas behind CTC based ASR were first introduced in the following papers \cite{graves2006connectionist,graves2014towards}. In our work we used a single layer gated recurrent unit (GRU) \cite{chung2014empirical} with 128 hidden units as encoder for the CTC network. The decoder consists of a combination of a dense layer ( fully connected layer) and a softmax activation. Output at every time step of the GRU layer is fed into the decoder network. 
The number of time steps of the GRU encoder is equal to product of the sampling frequency of the input features and the length of the input sequence. Since different speakers have different rate of speech, we used dynamic recurrent neural network (RNN) cell. There is no fixed value for time steps of the encoder.\\
Usually the number of time steps of the encoder (\textbf{T}) is greater than the length of output tokens for a continuous speech recognition problem. A RNN based CTC network tries to make length of output tokens equal to \textbf{T} by allowing the repetition of output prediction unit tokens and by introducing a special token called blank token \cite{graves2006connectionist} across all the frames. 
We used CTC loss function with adam optimizer \cite{kingma2014adam} and during inference time we used CTC beam search decoder. 

We now explain the loss function used in our CTC model. Consider training data set $X$ with training examples $\vec{x_1}, \cdots, \vec{x_m}$ and the corresponding label set $Y$ with target vectors $\vec{y_1}, \cdots, \vec{y_m}$. Consider any training example, label pair ($x$,$y$). 
Let the number of time steps of the RNN encoder for ($x$,$y$) is $T$. In case of character based CTC model, the RNN predicts a character at every time step. Whereas in word based CTC model, the RNN predicts a word at every time step. For the sake of simplicity, let us assume that length of target vector $y$ is equal to $T$. Let the probability vector output by the RNN at each time step $t$ be $\overrightarrow{z_t}$ and let $k^{th}$ value of $z_t$ be denoted by $z_t[k]$. The probability that model outputs $y$ on input $x$ is given by $Pr(y|x) = \prod_{t = 1}^T z_t[y[t]]$. During the training phase, we would like to maximize the conditional probability $Pr(y|x)$, and thereby define the loss function as $-\log Pr(y|x)$.

 In case when the length of $y$ is less than $T$, we extend the target vector $y$ by repeating a few of its values and by introducing blank token ($\epsilon$) to create a target vector of length $T$. Let the possible extensions of $y$ be denoted by $a_1, a_2, \cdots, a_\ell$. For example, when $y = [c,u,t]$ and $T = 4$, the possible extensions are $a_1 = [c,c,u,t]$, $a_2 = [c,u,u,t]$, $a_3 = [c,u,t,t]$, $a_4 = [\epsilon,c,u,t]$,
$a_5 = [c,\epsilon,u,t]$, $a_6 = [c,u,\epsilon,t]$ and $a_7 = [c,u,t,\epsilon]$. We then define $Pr(y|x)$ as $\sum_{i=1}^\ell Pr(a_i|x)$.\\
In our work we used character based CTC ASR model.
CTC assumes the conditional independence constraint that output predictions are independent given the entire input sequence.
\subsection{RNN Encoder-Decoder or Attention model}
RNN encoder - decoder ASR model consists of a RNN encoder and a RNN decoder with attention mechanism \cite{cho2014learning,chorowski2015attention,bahdanau2014neural}. The number of time steps of the encoder is equal to the product of sampling frequency of the input features and the length of input sequence. There is no fixed value for time steps in our case. We used dynamic RNN cell. We used a single layer GRU with 128 hidden units for both encoder and decoder. A dense layer followed by softmax activation is used after the decoder GRU to get the prediction probabilities. Dense layer performs an affine transformation.  The number of time steps of the decoder GRU is same as the number of words present in the sentence for a given training example. Training objective is to maximize the log probability of the ordered conditionals, ie: $\log Pr(Y_1|X) + \log Pr(Y_2|X,Y_1) + \log Pr(Y_3|X, Y_1,Y_2) \cdots +\log Pr(Y_u|X,Y_1,\cdots \cdots,Y_u) $,  
where X is input feature vector, $Y_i$\textquotesingle s are the labels for the  ordered words present in that training example and $u$ is the length of the output label sentence for that example. Cross entropy was used as the loss function with adam as the optimizer. We used teacher forcing algorithm \cite{williams1989learning} to train the model. 
During inference time we used beam search decoder. \\
We now explain the attention mechanism used in our attention model. Consider any training example, label pair ($x$,$y$). Let the number of times steps of encoder GRU for that example be $T$. The GRU encoder will transform the input features (${x_1}, {x_2}, \cdots \cdots ,{x_T}$) into hidden output feature vectors ($\vec{h_1}, \vec{h_2}, \cdots \cdots ,\vec{h_T}$). Let $k^{th}$ word label in $\vec{y}$ (sentence) be $y_k$, then to predict $y_k$ at decoder time step $k$, context vector $c_k$ is computed and fed into the decoder GRU. $c_k$ is computed as $\sum_{t=1}^{T}\vec{h_t}\alpha_{k,t}$ , where $\alpha_{k,t} $ is the attention weight vector satisfying the property $\sum_{t=1}^{T}\alpha_k,_t=1$.\\
$\alpha_{k,t}$ can be intuitively seen as a measure of how much attention $y_k$ must pay to $\vec{h_t}$, $t=\{1, 2, 3,\cdots \cdots,T\}$.   
$\alpha_{k,t}$ is mathematically defined as $softmax(score(\vec{h_t},\vec{h_{s-1}}))$, where $\vec{h_{s-1}}$ is hidden state of the decoder GRU at time step $k-1$.\\
The way of computing value for $score(\vec{h_t},\vec{h_{s-1}})$ depends on the type of attention used. In this work, we used bahdanau's additive style attention \cite{chorowski2015attention}, which defines $score(\vec{h_t},\vec{h_{s-1}})$ as 
$V\cdot tanh(W_1 \cdot \vec{h_t} + W_2 \cdot \vec{h_{s-1}}$) where $V, W_1$ and $W_2$ are learnable parameters during training of the model. 
\begin{figure}[h]
\begin{center}
\includegraphics[height=3cm,width=0.25\textwidth,trim={1cm 1cm 1cm 0.1cm},clip]{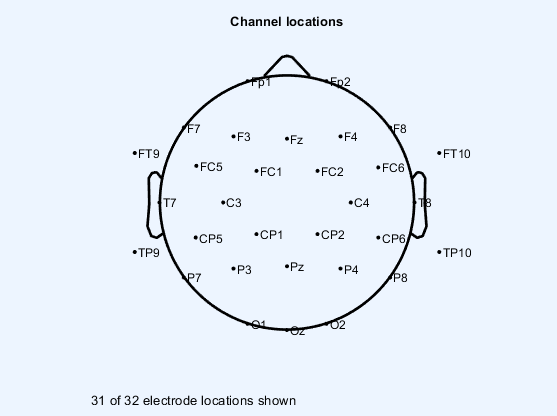}
\caption{EEG channel locations for the cap used in our experiments} 
\label{1vsall}
\end{center}
\end{figure}

\section{Design of Experiments for building the database}
We built two types of simultaneous speech EEG recording databases for this work. For database A five female and five male subjects took part in the experiment. For database B five male and three female subjects took part in the experiment. Except two subjects, rest all were native English speakers for both the databases. All subjects were UT Austin undergraduate,graduate students in their early twenties.\\
For data set A, the 10 subjects were asked to speak the first 30 sentences from the USC-TIMIT database\cite{narayanan2014real} and their simultaneous speech and EEG signals were recorded. This data was recorded in presence of background noise of 40 dB (noise generated by room air conditioner fan). We then asked each subject to repeat the same experiment two more times, thus we had 30 speech EEG recording examples for each sentence.\\
For data set B, the 8 subjects were asked to repeat the same previous experiment but this time we used background music played from our lab computer to generate a background noise of 65 dB. Here we had 24 speech EEG recording examples for each sentence.\\
We used Brain Vision EEG recording hardware. Our EEG cap had 32 wet EEG electrodes including one electrode as ground as shown in Figure 1. We used EEGLab \cite{delorme2004eeglab} to obtain the EEG sensor location mapping. It is based on standard 10-20 EEG sensor placement method for 32 electrodes.\\ 
For data set A, we used data from first 8 subjects for training the model, remaining two subjects data for validation and test set respectively.\\
For data set B, we used data from first 6 subjects for training the model, remaining two subjects data for validation and test set respectively. 
\section{EEG and Speech feature extraction details}
EEG signals were sampled at 1000Hz and a fourth order IIR band pass filter with cut off frequencies 0.1Hz and 70Hz was applied. A notch filter with cut off frequency 60 Hz was used to remove the power line noise.
EEGlab's \cite{delorme2004eeglab} Independent component analysis (ICA) toolbox was used to remove other biological signal artifacts like electrocardiography (ECG), electromyography (EMG), electrooculography (EOG) etc from the EEG signals. 
We extracted five statistical features for EEG, namely root mean square, zero crossing rate,moving window average,kurtosis and power spectral entropy \cite{krishna2019speech}. So in total we extracted 31(channels) X 5 or 155 features for EEG signals.The EEG features were extracted at a sampling frequency of 100Hz for each EEG channel.\\
We used spectral entropy because it captures the spectral ( frequency domain) and signal complexity information of EEG. It is also a widely used feature in EEG signal analysis\cite{zhang2008feature}. Similarly zero crossing rate was chosen as it is a commonly used feature both for speech recognition and bio signal analysis.
Remaining features were chosen to capture time domain statistical information. We performed lot of experiments to identify this set of features. Initially we used only spectral entropy and zero crossing rate but we noticed that the performance of the ASR system went up when we added the remaining additional features.\\ 
The recorded speech signal was sampled at 16KHz frequency. We extracted Mel-frequency cepstrum coefficients (MFCC) as features for speech signal.
We first extracted MFCC 13 features and then computed first and second order differentials (delta and delta-delta) thus having total MFCC 39 features.
The MFCC features were also sampled at 100Hz same as the sampling frequency of EEG features to avoid seq2seq problem.
\begin{figure}[h]
\centering
\includegraphics[height=5cm, width=0.4
\textwidth,trim={0.1cm 0.1cm 0.1cm 0.1cm},clip]{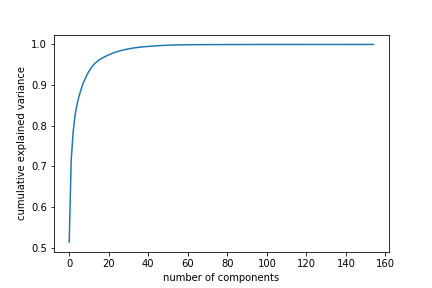}
\caption{Explained variance plot}
\label{1vsall}
\end{figure}
\section{EEG Feature Dimension Reduction Algorithm Details}
After extracting EEG and acoustic features as explained in the previous section, we used non linear methods to do feature dimension reduction in order to obtain set of EEG features which are better representation of acoustic features. 
We reduced the 155 EEG features to a dimension of 30 by applying Kernel Principle Component Analysis (KPCA) \cite{mika1999kernel}.We plotted cumulative explained variance versus number of components to identify the right feature dimension as shown in Figure 2. We used KPCA with polynomial kernel of degree 3 \cite{krishna2019speech}. 
We further computed delta, delta and delta of those 30 EEG features, thus the final feature dimension of EEG was 90 (30 times 3) for both the data sets.\\
The non linear dimension reduction of EEG features helped in improving the performance of ASR.
\section{Results}
The attention model was predicting a word and CTC model was predicting a character at every time step, hence we used word error rate (WER) as performance metric to evaluate attention model and character error rate (CER) for CTC model for different feature sets as shown below.\\
Table \rom{1} and \rom{2} shows the test time results for attention model for both the data sets when trained using EEG features and concatenation of EEG, acoustic features respectively. As seen from the results the attention model gave lower WER when trained and tested on smaller number of sentences. As the vocabulary size increase, the WER also went up. We believe for the attention model to achieve lower WER for larger vocabulary size more number of training examples or larger training data set is required as large number of weights need to be adapted. Figure 3 shows the training loss convergence of our attention model.\\
Table \rom{4} and \rom{5} shows the results obtained using CTC model. The error rates for CTC model also went up with the increase in vocabulary size for both the data sets. 
However the CTC model was trained for 500 epochs compared to 100
epochs for attention model to observe loss convergence and batch size was set to one for CTC model. Thus CTC model training was lot more time consuming than attention model.

In \cite{krishna2019speech} we have demonstrated that EEG sensors T7 and T8 features contributed most towards ASR performance. Table \rom{6} shows the CTC model test time results when we trained the model using EEG features from only T7 and T8 sensors on the most noisy data set B. We observed that as vocabulary size increase, error rates were slightly lower than the error rates from Table \rom{4} where we used EEG features from all 31 sensors with dimension reduction. Table \rom{3} shows the results for attention model when trained with EEG features from sensors T7 and T8 only on data set B. We  observed that error rates were higher in this case compared to the error rates reported in table \rom{2}.\\ 
Figures 4 shows the visualization of the attention weights when the attention model was trained and tested using only EEG features for Data set B. The plots shows the EEG feature importance ( attention) distribution across time steps for predicting first sentence and it indicates that attention model was not able to attend properly to EEG features, which might be another reason for giving higher WER. 

\begin{figure}[h]
\includegraphics[height=4.5cm, width=0.4
\textwidth,trim={0.1cm 0.1cm 0.1cm 0.1cm},clip]{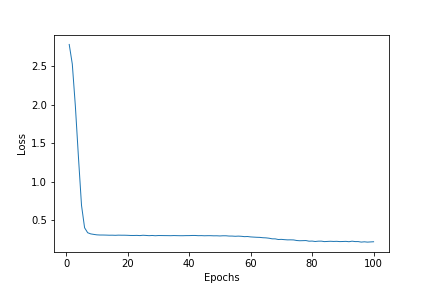}
\caption{Training loss convergence for attention model using only EEG features for first 10 sentences from data set A}
\label{1vsall}
\end{figure}
\begin{figure}[h]
\centering
\includegraphics[height=4.5cm, width=0.4
\textwidth,trim={0.1cm 0.1cm 0.1cm 0.1cm},clip]{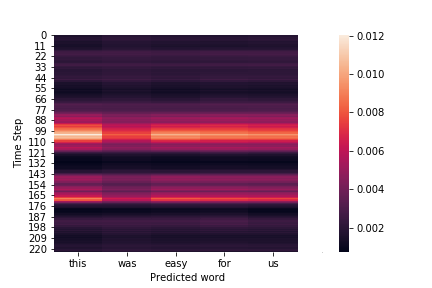}
\caption{Visualization of attention weights for the first sentence}
\label{1vsall}
\end{figure}
\begin{figure}[h]
\includegraphics[height=5cm, width=0.4
\textwidth,trim={0.1cm 0.1cm 0.1cm 0.1cm},clip]{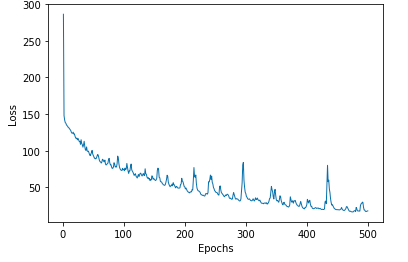}
\caption{Training loss convergence for CTC model using only EEG features for first 3 sentences from data set B}
\label{1vsall}
\end{figure}
\begin{table}[!ht]
\centering
\begin{tabular}{|l|l|l|l|}
\hline
\textbf{\begin{tabular}[c]{@{}l@{}}Number\\ of \\ Sentences\end{tabular}} & \textbf{\begin{tabular}[c]{@{}l@{}}Number of\\ unique\\ words\\ contained\end{tabular}} & \multicolumn{1}{c|}{\textbf{\begin{tabular}[c]{@{}c@{}}EEG\\ (WER \%)\end{tabular}}} & \multicolumn{1}{c|}{\textbf{\begin{tabular}[c]{@{}c@{}}EEG\\ +\\ MFCC\\ (WER \%)\end{tabular}}} \\ \hline
3                                                                         & 19                                                                                      & 0                                                                                    & \textbf{0}                                                                                      \\ \hline
5                                                                         & 29                                                                                      & \textbf{37.03}                                                                       & 45.1                                                                                            \\ \hline
7                                                                         & 42                                                                                      & 58.9                                                                                 & \textbf{51.1}                                                                                   \\ \hline
10                                                                        & 59                                                                                      & \textbf{63.11}                                                                       & 71.2                                                                                            \\ \hline
15                                                                        & 84                                                                                      & 82.7                                                                                 & \textbf{79}                                                                                     \\ \hline
20                                                                        & 106                                                                                     & 87.3                                                                                 & \textbf{80}                                                                                     \\ \hline
\end{tabular}
\caption{WER on test set for \textbf{attention model} for \textbf{Data set A}}
\end{table}
\begin{table}[!ht]
\centering
\begin{tabular}{|l|l|l|l|}
\hline
\textbf{\begin{tabular}[c]{@{}l@{}}Number\\ of \\ Sentences\end{tabular}} & \textbf{\begin{tabular}[c]{@{}l@{}}Number of\\ unique\\ words\\ contained\end{tabular}} & \multicolumn{1}{c|}{\textbf{\begin{tabular}[c]{@{}c@{}}EEG\\ (WER \%)\end{tabular}}} & \multicolumn{1}{c|}{\textbf{\begin{tabular}[c]{@{}c@{}}EEG\\ +\\ MFCC\\ (WER \%)\end{tabular}}} \\ \hline
3                                                                         & 19                                                                                      & 0                                                                                    & \textbf{0}                                                                                      \\ \hline
5                                                                         & 29                                                                                      & 44.4                                                                                 & \textbf{41.9}                                                                                   \\ \hline
7                                                                         & 42                                                                                      & \textbf{52.8}                                                                        & 55                                                                                              \\ \hline
10                                                                        & 59                                                                                      & \textbf{68}                                                                          & 71                                                                                              \\ \hline
15                                                                        & 84                                                                                      & \textbf{82.7}                                                                        & 83                                                                                              \\ \hline
20                                                                        & 106                                                                                     & 86.4                                                                                 & \textbf{86}                                                                                     \\ \hline
\end{tabular}
\caption{WER on test set for \textbf{attention model} for \textbf{Data set B}}
\end{table}
\begin{table}[!ht]
\centering
\begin{tabular}{|l|l|l|}
\hline
\textbf{\begin{tabular}[c]{@{}l@{}}Number\\ of Sentences\end{tabular}} & \textbf{\begin{tabular}[c]{@{}l@{}}Number of \\ unique \\ words\\ contained\end{tabular}} & \textbf{\begin{tabular}[c]{@{}l@{}}EEG\\ (WER \%)\end{tabular}} \\ \hline
3                                                                      & 19                                                                                        & {\color[HTML]{333333} 35.2}                                     \\ \hline
5                                                                      & 29                                                                                        & {\color[HTML]{333333} 59.2}                                     \\ \hline
7                                                                      & 42                                                                                        & {\color[HTML]{333333} 71.7}                                     \\ \hline
10                                                                     & 59                                                                                        & {\color[HTML]{333333} 77.1}                                     \\ \hline
\end{tabular}
\caption{WER on test set for \textbf{attention model} for \textbf{Data set B} using EEG features from only \textbf{T7} and \textbf{T8} electrodes}
\end{table}
\begin{table}[!ht]
\centering
\begin{tabular}{|l|l|l|l|}
\hline
\textbf{\begin{tabular}[c]{@{}l@{}}Number \\ of\\ Sentences\end{tabular}} & \textbf{\begin{tabular}[c]{@{}l@{}}Number of\\ unique\\ words\\ contained\end{tabular}} & \textbf{\begin{tabular}[c]{@{}l@{}}EEG\\ (CER\%)\end{tabular}} & \multicolumn{1}{c|}{\textbf{\begin{tabular}[c]{@{}c@{}}EEG\\ +\\ MFCC\\ (CER\%)\end{tabular}}} \\ \hline
3                                                                         & 19                                                                                      & \textbf{26.3}                                                  & 53                                                                                             \\ \hline
5                                                                         & 29                                                                                      & 68.7                                                           & \textbf{61.2}                                                                                  \\ \hline
7                                                                         & 42                                                                                      & 66                                                             & \textbf{64.2}                                                                                  \\ \hline
10                                                                        & 59                                                                                      & 67.2                                                           & \textbf{66}                                                                                    \\ \hline
15                                                                        & 84                                                                                      & 74.3                                                           & \textbf{68.2}                                                                                  \\ \hline
20                                                                        & 106                                                                                     & 77.5                                                           & \textbf{67.9}                                                                                  \\ \hline
\end{tabular}
\caption{CER on test set for \textbf{CTC model} for \textbf{Data set B}}
\end{table}
\begin{table}[!ht]
\centering
\begin{tabular}{|l|l|l|l|}
\hline
\textbf{\begin{tabular}[c]{@{}l@{}}Number \\ of\\ Sentences\end{tabular}} & \textbf{\begin{tabular}[c]{@{}l@{}}Number of\\ unique\\ words\\ contained\end{tabular}} & \textbf{\begin{tabular}[c]{@{}l@{}}EEG\\ (CER\%)\end{tabular}} & \multicolumn{1}{c|}{\textbf{\begin{tabular}[c]{@{}c@{}}EEG\\ +\\ MFCC\\ (CER\%)\end{tabular}}} \\ \hline
3                                                                         & 19                                                                                      & 45.8                                                           & \textbf{10.27}                                                                                 \\ \hline
5                                                                         & 29                                                                                      & \textbf{45}                                                    & 47                                                                                             \\ \hline
7                                                                         & 42                                                                                      & \textbf{61.02}                                                 & 63                                                                                             \\ \hline
10                                                                        & 59                                                                                      & 70                                                             & \textbf{61}                                                                                    \\ \hline
15                                                                        & 84                                                                                      & 75.4                                                           & \textbf{64.3}                                                                                  \\ \hline
20                                                                        & 106                                                                                     & 73.91                                                          & \textbf{67.09}                                                                                 \\ \hline
\end{tabular}
\caption{CER on test set for \textbf{CTC model} for \textbf{Data set A}}
\end{table}
\begin{table}[!ht]
\centering
\begin{tabular}{|l|l|l|}
\hline
\textbf{\begin{tabular}[c]{@{}l@{}}Number\\ of \\ Sentences\end{tabular}} & \textbf{\begin{tabular}[c]{@{}l@{}}Number of\\ unique\\ words\\ contained\end{tabular}} & \multicolumn{1}{c|}{\textbf{\begin{tabular}[c]{@{}c@{}}EEG\\ (CER \%)\end{tabular}}} \\ \hline
3                                                                         & 19                                                                                      & 40.5                                                                                 \\ \hline
5                                                                         & 29                                                                                      & 60.3                                                                                 \\ \hline
7                                                                         & 42                                                                                      & 65                                                                                   \\ \hline
10                                                                        & 59                                                                                      & 65.7                                                                                 \\ \hline
\end{tabular}
\caption{CER on test set for \textbf{CTC model} for \textbf{Data set B} using EEG features from only \textbf{T7} and \textbf{T8} electrodes}
\end{table}

\begin{table}[!ht]
\centering
\begin{tabular}{|l|l|l|l|}
\hline
\textbf{\begin{tabular}[c]{@{}l@{}}Number \\ of\\ Sentences\end{tabular}} & \textbf{\begin{tabular}[c]{@{}l@{}}Number of\\ unique\\ words\\ contained\end{tabular}} & \textbf{\begin{tabular}[c]{@{}l@{}}MFCC\\ (CER\%)\end{tabular}} & \multicolumn{1}{c|}{\textbf{\begin{tabular}[c]{@{}c@{}}EEG\\ +\\ MFCC\\ (CER\%)\end{tabular}}} \\ \hline
15                                                                        & 84                                                                                      & 65.42                                                           & \textbf{64.3}                                                                                  \\ \hline
20                                                                        & 106                                                                                     & 67.85                                                           & \textbf{67.09}                                                                                 \\ \hline
\end{tabular}
\caption{CER on test set for \textbf{CTC model} for \textbf{Data set A} for \textbf{MFCC-EEG fusion} for larger vocabulary size}
\end{table}

\begin{table}[!ht]
\centering
\begin{tabular}{|l|l|l|l|}
\hline
\textbf{\begin{tabular}[c]{@{}l@{}}Number \\ of\\ Sentences\end{tabular}} & \textbf{\begin{tabular}[c]{@{}l@{}}Number of\\ unique\\ words\\ contained\end{tabular}} & \textbf{\begin{tabular}[c]{@{}l@{}}MFCC\\ (CER\%)\end{tabular}} & \multicolumn{1}{c|}{\textbf{\begin{tabular}[c]{@{}c@{}}EEG\\ +\\ MFCC\\ (CER\%)\end{tabular}}} \\ \hline
15                                                                        & 84                                                                                      & \textbf{65.7}                                                   & 68.2                                                                                           \\ \hline
20                                                                        & 106                                                                                     & \textbf{67.9}                                                     & \textbf{67.9}                                                                                    \\ \hline
\end{tabular}
\caption{CER on test set for \textbf{CTC model} for \textbf{Data set B} for \textbf{MFCC-EEG fusion} for larger vocabulary size}
\end{table}
\section{Conclusion and Future work}
In this paper we demonstrated the feasibility of using EEG features, concatenation of EEG and acoustic features for performing noisy continuous speech recognition. To our best knowledge this is the first time a continuous noisy speech recognition is demonstrated using only EEG features.\\
For both attention and CTC model we observed that as the vocabulary size increase, concatenating acoustic features with EEG features will help in reducing the test time error rates.\\ 
We further plan to publish our speech EEG data base used in this work to help advancement of research in this area.\\
For future work, we plan to build a much larger speech EEG data base and also perform experiments with data collected from subjects with speaking disabilities.\\
We will also investigate whether it is possible to improve the attention model results by tuning hyper parameters to improve the model's ability to condition on the input,improve CTC model results by training with more number of examples and by using external language model during inference time.

\section{Acknowledgement} 
We would like to thank Kerry Loader and Rezwanul Kabir from Dell, Austin, TX for donating us the GPU to train the models used in this work.\\

\bibliographystyle{IEEEtran}

\bibliography{refs}
\end{document}